\documentclass[english,aps,reprint,superscriptaddress,twocolumn]{revtex4-2}
\usepackage[T1]{fontenc}
\usepackage[latin9]{inputenc}
\usepackage{units}
\usepackage{bm}
\usepackage{amsmath}
\usepackage{amssymb}
\usepackage{graphicx}
\usepackage{esint}

\makeatletter
\usepackage[english]{babel}
\usepackage{graphicx}
\usepackage{dcolumn}
\usepackage{bm}
\usepackage{xcolor}
\def\urlprefix{}
\def\url#1{}
\addto\captionsenglish{%
}

\usepackage{babel}
\usepackage[none]{hyphenat}

\newcommand{\fgt}{$\text{Fe}_3\text{GeTe}_2$}

\newcommand{\fref}[1]{Fig.~\ref{#1}}

\providecommand*{\diff}%
{\@ifnextchar^{\DIfF}{\DIfF^{}}}
\def\DIfF^#1{%
	\mathop{\mathrm{\mathstrut d}}%
	\nolimits^{#1}\gobblespace}
\def\gobblespace{%
	\futurelet\diffarg\opspace}
\def\opspace{%
	\let\DiffSpace\!%
	\ifx\diffarg(%
	\let\DiffSpace\relax
	\else
	\ifx\diffarg[%
	\let\DiffSpace\relax
	\else
	\ifx\diffarg\{%
	\let\DiffSpace\relax
	\fi\fi\fi\DiffSpace}
\usepackage{babel}

\makeatother

\usepackage{babel}
\begin{document}
\title{Spin scattering and Hall effects in monolayer $\textrm{Fe}_{3}\textrm{Ge}\textrm{Te}_{2}$}
\author{Luyan Yu}
\affiliation{Department of Physics, University of Texas at Austin, Austin, Texas
78712, USA.}
\author{Jie-Xiang Yu}
\affiliation{Department of Physics, University of Florida, Gainesville, Florida
32611, USA}
\author{Jiadong Zang}
\affiliation{Department of Physics and Materials Science Program, University of
New Hampshire, Durham, NH 03824, USA}
\author{Roger K. Lake}
\affiliation{Laboratory for Terahertz \& Terascale Electronics (LATTE), Department
of Electrical and Computer Engineering, University of California-Riverside,
Riverside, California 92521, USA}
\author{Houlong Zhuang}
\affiliation{Mechanical and Aerospace Engineering, Arizona State University, Tempe,
Arizona 85287, USA}
\author{Gen Yin}
\thanks{gen.yin@georgetown.edu}
\affiliation{Department of Physics, Georgetown University, Washington, D.C. 20057,
USA}
\begin{abstract}
We theoretically show that the carrier transport in monolayer \fgt\
experiences a transition between anomalous Hall effect and spin Hall
effect when the spin polarization of disorders switches between out-of-plane
and in-plane. These Hall effects are allowed when the magnetization
is polarized in-plane, breaking the ${\cal C}_{3}$ rotation symmetry.
The transition originates from the selection rule of spin scattering,
the strong spin-orbit coupling, and the van Hove singularities near
the Fermi surface. The scattering selection rule tolerates the sign
change of the disorder spin, which provides a convenient method to
detect the switching of antiferromagnetic insulators regardless of
the interfacial roughness in a heterostructure. This provides a convenient
platform for the study of 2D spintronics through various van-der-Waals
heterostructures. 
\end{abstract}
\maketitle

As a two-dimensional (2D) magnet, \fgt\ (FGT) has a surprisingly
robust long-range ferromagnetic order with a perpendicular easy axis
and a reasonably high Curie temperature\citep{deiseroth_fe3gete2_2006,chen_magnetic_2013,zhuang_strong_2016,fei_two-dimensional_2018,deng_gate-tunable_2018,tan_hard_2018,may_ferromagnetism_2019}.
Distinct from many 2D spintronic materials discovered recently \citep{gong_discovery_2017,huang_electrical_2018},
the family of $\textrm{Fe}_{3,5}\textrm{Ge}\textrm{Te}_{2}$ are known
as Ising itinerant 2D magnets, owing to their unique gapless spectrum
and the sizable perpendicular anisotropy\citep{ribeiro_large-scale_2022,zhu_electronic_2016,yi_competing_2016}.
Such anisotropy is a consequence of the strong spin-orbit coupling
(SOC) given by the Te atoms, which also strongly impacts the transport
behavior of carriers, resulting in a sizable anomalous Hall effect
\citep{kim_large_2018,deng_gate-tunable_2018,tan_hard_2018}. Without surface dangling
bonds, few-layer FGTs can provide atomically sharp interfaces, resulting
in high-quality heterostructures. The vast parameter space of stacking
and twisting also enables the modulation of the transport and magnetic
properties in a large range\citep{gong_two-dimensional_2019,gibertini_magnetic_2019,wang_tunneling_2018,wu_ne-type_2020,wang_current-driven_2019,huang_emergent_2020}.
These advantages make FGT an intriguing platform to investigate 2D
magnetism as well as to implement next-generation low-dimensional
spintronic devices. 
%

Although the anomalous Hall effect in bulk FGT is sizable, 
it is expected to be small in monolayers, which motivates us
to investigate extrinsic Hall effects induced by disorder.
For bulk FGT, a symmetry-protected nodal line results in a large
local Berry curvature and a large intrinsic anomalous Hall effect\citep{kim_large_2018}.
However, such nodal line is along the direction perpendicular to the
vdW planes, which vanishes in the case of a monolayer\citep{lin_layer-dependent_2019}.
Furthermore, when forming spintronic interfaces with non-vdW materials,
the itinerant carriers can scatter from the disordered interface.
The interface disorder can be spin-polarized and respond to
an external magnetic field. 
The spin of the disordered interfacial atoms can also be
pinned when the local atomic orbitals are closely coupled to an adjacent
layer of magnetic or antiferromagnetic insulator with a higher ordering
temperature. 
The interface scattering can be particularly important when
the surface roughness of the adjacent magnetic layer destroys the long-range
order at the interface. 
Moreover, due to the strong SOC inherent to Te, the
spin texture, the symmetry, and the geometry of the Fermi surface are
sensitive to the direction of magnetization. 
This leads to intricate
selection rules of spin-dependent scattering and transport. 
Here, we show that the spin-dependent scattering in monolayer FGT works
together with the van Hove singularities near the Fermi surface, resulting
in a switching between an anomalous Hall effect and a spin Hall effect.
These Hall effects are allowed by an in-plane magnetization that breaks
the atomic ${\cal C}_{3}$ rotation symmetry, which otherwise forbids
any leading-order Hall effects for all individual bands. 
These transport
signatures can provide information of the spin-polarized disorders
when a monolayer of FGT is weakly coupled to an insulating magnetic
or antiferromagnetic system, providing a convenient experimental probe
of the switching through carrier transport. 

\begin{figure}
\begin{centering}
\includegraphics[width=1\columnwidth]{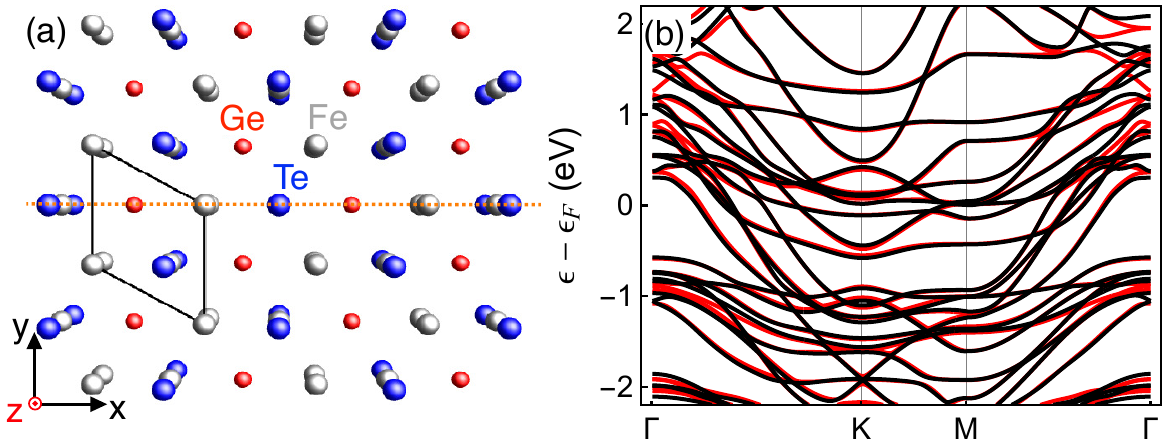}
\par\end{centering}
\begin{centering}
\includegraphics[width=1\columnwidth]{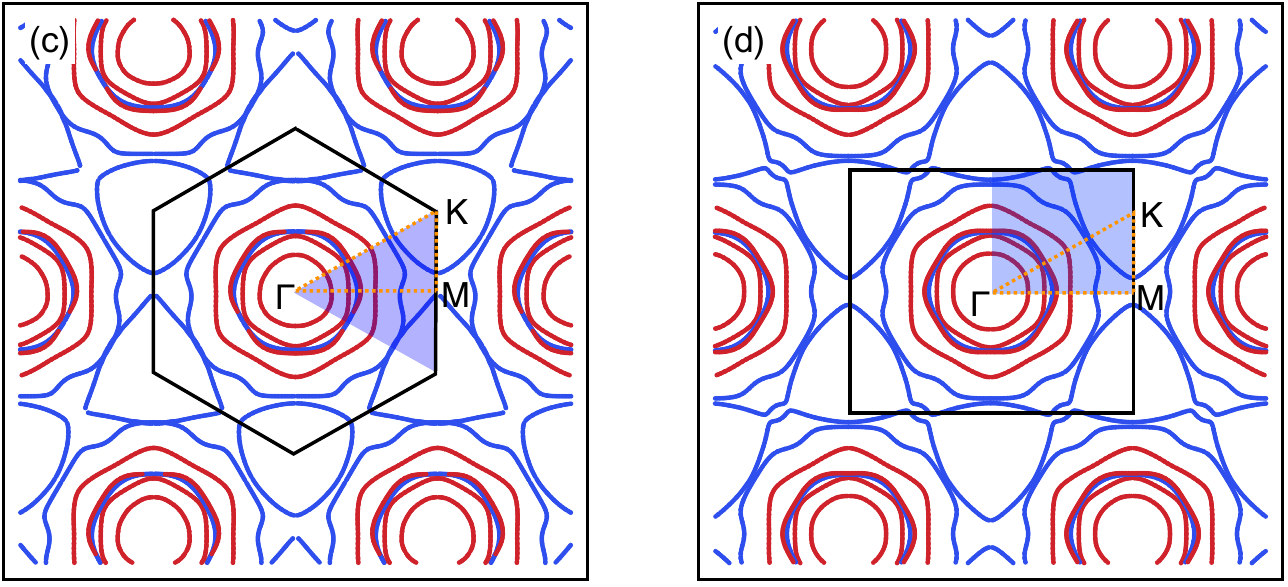}
\par\end{centering}
\centering{}\caption{(a) The top view of a monolayer FGT. The unit cell is indicated by
the rhombic prism containing three Fe atoms (gray), 1 Ge atom (red)
and 2 Te atoms (blue). (b) Band structure of monolayer FGT along high
symmetry points with spins polarized along $\hat{z}$ (dark) and $\hat{x}$
(red). (c-d) The Brillouin zones and the smallest repeating wedges
(blue transparent) when the magnetization is along $\hat{z}$ (c)
and $\hat{x}$ (d). Positive and negative spins are denoted by the
red and blue solid lines, respectively. }
\label{fig:gMaterial}
\end{figure}

The atomic structure of a monolayer FGT is illustrated in Fig.\ref{fig:gMaterial}(a),
where a top view is presented with some graphic perspective to show
the vertical alignment of the atoms. 
The spectrum of the material
is obtained from first-principles calculations within the framework
of density functional theory (DFT) using the projector augmented wave
pseudopotential \citep{blochl_projector_1994,kresse_ultrasoft_1999}
as implemented in Vienna \emph{Ab initio} Simulation Package (VASP)
\citep{kresse_efficiency_1996,kresse_efficient_1996}. 
The local density
approximation \citep{perdew_self-interaction_1981} was used for the
exchange-correlation energy. 
A $600\thinspace\textrm{eV}$ energy
cutoff for the plane-wave expansion was used throughout the calculations.
The $\textrm{\ensuremath{\Gamma}}$-centered mesh of $15\times15\times1$
in the two-dimensional Brillouin zone (BZ) was adopted. 
$a=3.90\thinspace\textrm{\AA}$
was chosen as the in-plane lattice constant for hexagonal lattice.
After we obtained the eigenstates and eigenvalues, a unitary transformation
of Bloch waves was performed to construct the tight-binding Hamiltonian
in a Wannier function basis by using the maximally localized Wannier
functions (WF) method \citep{marzari_maximally_2012} implemented
in the Wannier90 package \citep{mostofi_updated_2014}. 
The WF-based
Hamiltonian has the same eigenvalues as those obtained by first-principles
calculations within $1.0\thinspace\textrm{eV}$ of the Fermi level.
The band structure of the tight-binding model is shown in \fref{fig:gMaterial}(b).
Here, the red solid curves represent the case where the magnetization
is along $\hat{x}$, whereas the dark solid ones correspond to the
out-of-plane magnetization along $\hat{z}$. 
The presence of SOC is evidently captured. 
%

To examine the impact of the magnetization for low-temperature transport,
we show the Fermi surfaces for both polarization cases in Figs. \ref{fig:gMaterial}(c-d).
When the magnetization is along $\hat{z}$ {[}Fig. \ref{fig:gMaterial}(c){]},
the Fermi surface possesses most symmetries of the non-magnetic
crystal structure,
including an apparent ${\cal C}_{3}$ rotation. 
Note that the atomic
structure also possesses reflection symmetry with respect to the $\textrm{x-z}$
plane (${\cal R}_{xz}$), denoted by the orange dotted line in Fig.
\ref{fig:gMaterial}(a), and the mirror plane containing the Ge atoms
(${\cal R}_{xy}$). 
When considering the magnetization along $\hat{z}$,
the ${\cal C}_{3}$ rotation and the ${\cal R}_{xy}$ reflection remain
symmetric, whereas ${\cal R}_{xz}$ is broken by the spin. 
The irreducible wedge of the 1st Brillouin zone is therefore an equilateral
triangle as shown in Fig. \ref{fig:gMaterial}(c). 
The hexagonal full
Brillouin zone can then be restored by ${\cal C}_{3}$ rotations and
another combination of reflection and time reversal (${\cal R}_{xy}{\cal T}$).
On the other hand, when the magnetization is along $\hat{x}$, the
spin breaks the ${\cal C}_{3}$ symmetry, which is evident in the
Fermi surface in Fig. \ref{fig:gMaterial}(d). 
In this case, only
${\cal R}_{xz}{\cal T}$ and ${\cal R}_{xy}{\cal T}$ remain symmetric.
The irreducible wedge of the Brillouin zone now becomes a rectangle
{[}blue in Fig. \ref{fig:gMaterial}(d){]}. 
The $\Gamma\rightarrow K\rightarrow M$
path used in Fig. \ref{fig:gMaterial}(b) is thus no longer a unique
high-symmetry path, as shown in Fig. \ref{fig:gMaterial}(d). 
%

The change in the symmetry of the Fermi surface has a profound impact
on the cryogenic magnetotransport property. 
At the limit of linear
response governed by the Boltzmann transport equation, 
the transport of charge and
spin is driven by the asymmetric part of the distribution function,
$f_{a}=f-f_{0}=\tau_{\mathbf{k}}e\left.\frac{\partial f_{0}}{\partial\epsilon}\right|_{\epsilon_{F}}\boldsymbol{{\cal E}}\cdot\mathbf{v}_{\mathbf{k}}$,
where $-e<0$ is the electron charge, $f_{0}$ is the equilibrium
distribution, $\mathbf{v}_{\mathbf{k}}$ is group velocity, and $\tau_{\mathbf{k}}$
is the $\mathbf{k}\textrm{-}$dependent relaxation time. 
At zero temperature,
the Hall and the longitudinal conductivities can be obtained by collecting
the Fermi-surface contributions from all bands, $\sigma_{\perp,\parallel}=\sum_{n}\sigma_{\perp,\parallel}^{(n)}$,
where $\sigma_{\perp,\parallel}^{(n)}=-\frac{e^{2}}{\hbar}\oint_{\epsilon=\epsilon_{F},}\tau_{n}\frac{v_{\parallel}^{(n)}v_{\perp,\parallel}^{(n)}}{v^{(n)}}dl$.
Here, $v_{\parallel}$ and $v_{\perp}$ represent the group-velocity
components that are parallel and perpendicular to $\boldsymbol{{\cal E}}$,
respectively. 
Within the first Brillouin zone, one can parameterize
the Fermi surfaces using the azimuthal angle of $\mathbf{k}=\mathbf{k}(\theta)$,
such that $\sigma_{\perp,n}=-\frac{e^{2}}{\hbar}\oint_{n}f(\theta)\cos\phi_{\theta}\sin\phi_{\theta}\diff\theta$,
where $\phi_{\theta}$ is the angle between $\mathbf{v}_{\theta}$
and $\boldsymbol{{\cal E}}$, and $f(\theta)=|\mathbf{v}_{\theta}|k_{\theta}\tau_{\theta}$.
In the case of out-of-plane magnetization, if all scattering mechanisms
preserve the crystalline symmetry, $f(\theta)$ should also have ${\cal C}_{3}$
rotation symmetry. 
Further using $\phi_{\theta+\frac{2\pi}{3}}=\phi_{\theta}+\frac{2\pi}{3}$,
we have
\begin{align}
\sigma_{\perp} & =-\frac{e^{2}}{\hbar}\int_{0}^{\frac{2\pi}{3}}f(\theta)\nonumber \\
 & \left[\cos\phi_{\theta}\sin\phi_{\theta}+\cos(\phi_{\theta}+\frac{2\pi}{3})\sin(\phi_{\theta}+\frac{2\pi}{3})\right.\nonumber \\
 & \left.+\cos(\phi_{\theta}+\frac{4\pi}{3})\sin(\phi_{\theta}+\frac{4\pi}{3})\right]\diff\theta=0.\label{eq:IntegralWithC3}
\end{align}
where Hall effects are strictly forbidden due to the zero integrand.
Note that the band index is omitted in Eq. \ref{eq:IntegralWithC3},
suggesting that Hall effects given by such Fermi surfaces are forbidden
for each individual band. 
However, when the magnetization is along
$\hat{x}$, the ${\cal C}_{3}$ symmetry of the Fermi surface is broken
due to the strong SOC. 
As a result the integrand in Eq. \ref{eq:IntegralWithC3}
becomes finite, allowing for electric or spin Hall effects.
We note that although FGT thin films are known to have a perpendicular easy axis, an in-plane saturation is experimentally feasible in a 4-layer thin film by applying an in-plane field of $\sim3\textrm{}T$ \cite{deng_gate-tunable_2018,tan_hard_2018}.

To mimic the scenario where a monolayer FGT is weakly coupled to a
spin system, we consider the scattering from a local spin:
\begin{equation}
\hat{H}=-J_{H}\Omega\mathbf{S}\cdot\hat{\mathbf{s}}\delta(\mathbf{r}),\label{eq:smat:e1}
\end{equation}
where $\mathbf{S}$ is a classical unit vector setting the polarization,
$\hat{\mathbf{s}}$ denotes Pauli matrices of the itinerant spin,
$J_{H}$ is the Hund's-rule exchange coupling, and $\Omega$ is the
area of the localized impurity. 
At the leading-order approximation,
the transition rate from $\mathbf{k}$ to $\mathbf{k}'$ is given
by Fermi's golden rule $S_{\mathbf{k}'\mathbf{k}}=\frac{2\pi N_{0}}{\hbar}|\langle\chi_{\mathbf{k}'}\rvert\hat{H}\lvert\chi_{\mathbf{k}}\rangle|^{2}\delta(\epsilon_{\mathbf{k}}-\epsilon_{\mathbf{k}'}),$
where $\lvert\chi_{\mathbf{k}}\rangle$ represents the periodic part
of the wave function at $\mathbf{k}$, and $N_{0}$ is the number
of impurity centers. 
Here, we consider $\mathbf{S}=\hat{x}$ and $\mathbf{S}=\hat{z}$
as two independent types of scattering mechanisms. 
When $\mathbf{S}=\hat{x}$,
the spin of the impurity is parallel to the magnetization. 
The Bloch
states therefore diagonalize the scattering Hamiltonian, allowing
only spin-preserved scattering: $|\langle x\pm|\hat{s}_{x}|x\pm\rangle|^{2}=1$.
However, when $S=\hat{z}$ or along any other direction within the
$y\textrm{-}z$ plane, spin-preserving scattering is forbidden: $|\langle x\pm|\hat{s}_{z}|x\pm\rangle|^{2}=0$,
allowing spin-flipping scatterings only. 
%

\begin{figure}
\centering{}\includegraphics[width=0.47\textwidth]{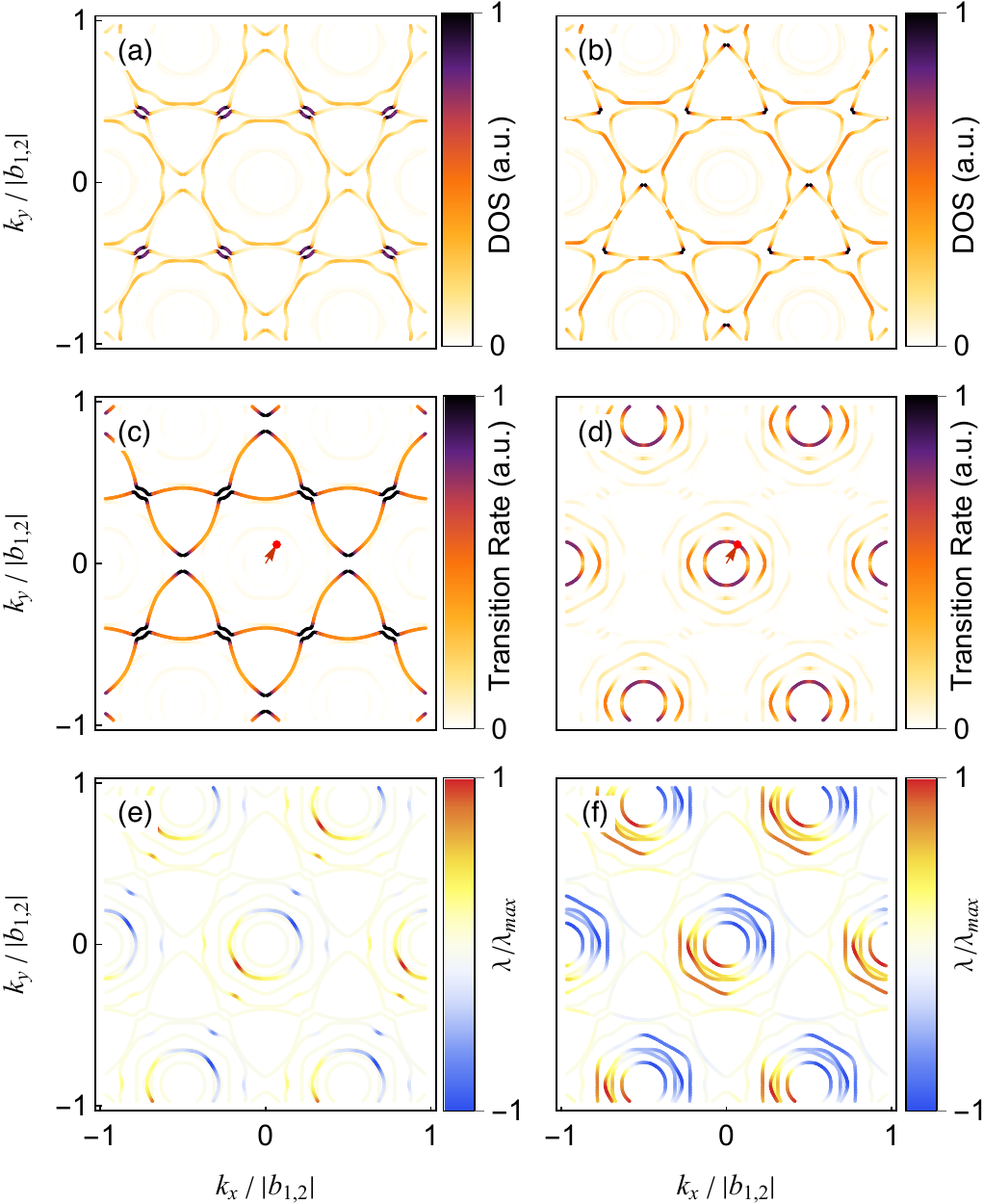} \caption{(a-b) The density of states (DOS) along the Fermi loops when the magnetization
is along $\hat{x}$ and $\hat{z}$, respectively. The color is normalized
to the maximum values of each plot. (c-d) The scattering for an initial
state denoted by the red solid dot highlighted by the dark-orange
arrow. The magnetization is along $\hat{x}$. The color denotes the
scattering rate to different final states on the Fermi space, normalized
to the maximum value. The scattering impurity is polarized along $\hat{z}$
and $\hat{x}$ for (c) and (d), respectively. (e-f) The free-flight displacements solved from full-band Boltzmann equation for the cases of $\mathbf{S}=\hat{z}$
and $\mathbf{S}=\hat{x}$, respectively. The color is normalized to
the maximum values. }
\label{fig:gPanel}
\end{figure}

The selection rules strongly impact the transport property of monolayer
FGT. 
This is a consequence of the spin composition and the density
of states (DOS) on the Fermi surface. When the magnetization is along
$\hat{x}$, four van Hove singularities {[}dark in Fig. \ref{fig:gPanel}(a){]}
are brought to the Fermi surface, resulting in large DOS for $|x-\rangle$.
However, these singularities vanish in the case of out-of-plane magnetization,
where the two Fermi loops enclosing $K$ and $K'$ are no longer intersecting
{[}Fig. \ref{fig:gPanel}(b){]}. The impact of these singularities
becomes apparent considering the spin selection rule. When the impurity
spins are along $\hat{z}$, the selection rule only allows the scattering
of $|x\pm\rangle\rightarrow|x\mp\rangle$. Due to the large DOS of
$|x-\rangle$, the scattering rate of $|x+\rangle\rightarrow|x-\rangle$
is dominating. In this case, $|x-\rangle$ has a much longer relaxation
time, and therefore dominates the transport. On the other hand, when
the impurities are polarized along $\hat{x}$, the selection rule
now allows only $|x\pm\rangle\rightarrow|x\pm\rangle$. The large
density of states for $|x-\rangle$ therefore makes the scattering
of $|x-\rangle\rightarrow|x-\rangle$ much more frequent. The relaxation
time for $|x+\rangle$ now becomes greater, thereby dominating the
transport and the Hall effects. Such transition of scattering rate
can be seen from Figs. \ref{fig:gPanel}(c-d), where we illustrate
the scattering rate from a chosen initial state (arrow, with the spin
of $|x+\rangle$) to all possible final states on the Fermi surface.
When $S=\hat{z}$, due to the spin selection rule, the initial positive
spin is only allowed to scatter to negative spins hosting the van
Hove singularities. This can be seen by the dark colors in Fig. \ref{fig:gPanel}(c).
In contrast, when the scattering centers are polarized along $\hat{x}$,
the originally dominating scattering is now forbidden, allowing only
same-spin scattering, as shown in Fig. \ref{fig:gPanel}(d). Note
that the spinless part of the Bloch states also affect the scattering
rate, resulting in the variation of transition rates even within the
same spin.

To quantitatively understand the transport signature of the spin-dependent
scattering centers, we evaluate the full-band collision integral.
Within the semi-classical approximation, the nonequilibrium distribution
function satisfies $-\frac{e\,\bm{\mathcal{E}}}{\hbar}\cdot\nabla_{\bm{k}}f=\left.\frac{\partial f}{\partial t}\right|_{\text{coll}}$.
Considering detailed balance, the collision integral is given by $\left.\frac{\partial f}{\partial t}\right|_{\text{coll}}=\sum_{\mathbf{k}'}S_{\mathbf{kk}'}f_{\mathbf{k}'}(1-f_{\mathbf{k}})-\sum_{\mathbf{k}'}S_{\mathbf{k}'\mathbf{k}}f_{\mathbf{k}}(1-f_{\mathbf{k}'})$,
where $S_{\mathbf{k}\mathbf{k}'}$ is the scattering rate from $\mathbf{k}'$
to $\mathbf{k}$. Using the leading-order approximation, we have $\left.\frac{\partial f}{\partial t}\right|_{\text{coll}}=-\frac{f_{a}}{\tau_{\mathbf{k}}}$.
Beyond the constant relaxation time approximation (RTA), $\tau_{\mathbf{k}}$
should be $\mathbf{k}\textrm{-}$dependent, satisfying $\boldsymbol{{\cal E}}\cdot\mathbf{v}_{\mathbf{k}}=\sum_{\mathbf{k}'}S_{\mathbf{k}'\mathbf{k}}\left(\tau_{\mathbf{k}}\boldsymbol{{\cal E}}\cdot\mathbf{v}_{\mathbf{k}}-\tau_{\mathbf{k}'}\boldsymbol{{\cal E}}\cdot\mathbf{v}_{\mathbf{k}'}\right)$,
where $\mathbf{v}_{\mathbf{k}}$ is the group velocity of the eigenstate
at $\mathbf{k}$. With some algebra, we have $v_{\mathbf{k}}^{\parallel}=-\sum_{\mathbf{k}'}S_{\mathbf{k}'\mathbf{k}}v_{\mathbf{k}'}^{\parallel}\tau_{\mathbf{k}'}+v_{\mathbf{k}}^{\parallel}\tau_{\mathbf{k}}\sum_{\mathbf{k}'}S_{\mathbf{k}'\mathbf{k}},$
where $v_{\mathbf{k}}^{\parallel}=\frac{\mathbf{v}_{\mathbf{k}}\cdot\boldsymbol{{\cal E}}}{{\cal E}}$.
Further define the free-flight displacement along the field $\lambda_{\mathbf{k}}^{\parallel}=v_{\mathbf{k}}^{\parallel}\tau_{\mathbf{k}}$,
the full-band Boltzmann equation reduces to a linear system: 
\begin{equation}
\left[\lambda^{\parallel}\right]=\left[\tilde{S}\right]^{\top}\left[\lambda^{\parallel}\right]+\left[\tilde{\lambda}^{\parallel}\right].\label{eq:boltzmann:e4}
\end{equation}
where $[\tilde{S}]_{\mathbf{k}'\mathbf{k}}$ and $[\tilde{\lambda}^{\parallel}]$
are the normalized scattering matrix and the vector of free-flight displacements,
respectively. Here, $[\tilde{S}]_{\mathbf{k}'\mathbf{k}}=S_{\mathbf{k}'\mathbf{k}}/\sum_{\mathbf{k}'}S_{\mathbf{k}'\mathbf{k}}$
and $[\tilde{\lambda}^{\parallel}]_{\mathbf{k}}=v_{\mathbf{k}}^{\parallel}/\sum_{\mathbf{k}'}S_{\mathbf{k}'\mathbf{k}}$.
Note that the linear system defined by Eq. \ref{eq:boltzmann:e4}
is general for arbitrary combinations of scattering mechanisms included
in $S_{\mathbf{k}'\mathbf{k}}$. Although in principle the matrices
are of uncountably infinite dimension, we will always deal with finite
ones due to discretization in practice. The choice of discretization
grid should be carefully made to avoid artificially breaking the symmetries
that the system intrinsically hosts. In this practice, an equilateral
triangular mesh is used to sample the irreducible wedge of the
Brillouin zone. To demonstrate the result, we rotate $\boldsymbol{{\cal E}}$
away from $\hat{x}$ by $50^{\circ}$, and the free-flight displacement $\lambda_{\mathbf{k}}^{\parallel}$
is plotted for $\mathbf{S}=\hat{z}$ and $\mathbf{S}=\hat{x}$ in
Figs. \ref{fig:gPanel}(e) and (f), where the corresponding spin distribution is illustrated in Fig. \ref{fig:gMaterial}(d). As discussed before,
the values of $\lambda_{\mathbf{k}}^{\parallel}$ for negative-spin bands are dominating when $\mathbf{S}=\hat{z}$,
as shown in Fig. \ref{fig:gPanel}(e). However, the $\lambda_{\mathbf{k}}^{\parallel}$
for $|x+\rangle$ become much greater when $\mathbf{S}=\hat{x}$,
as shown in Fig. \ref{fig:gPanel}(f).

\begin{figure}
\centering{}\includegraphics[width=0.5\textwidth]{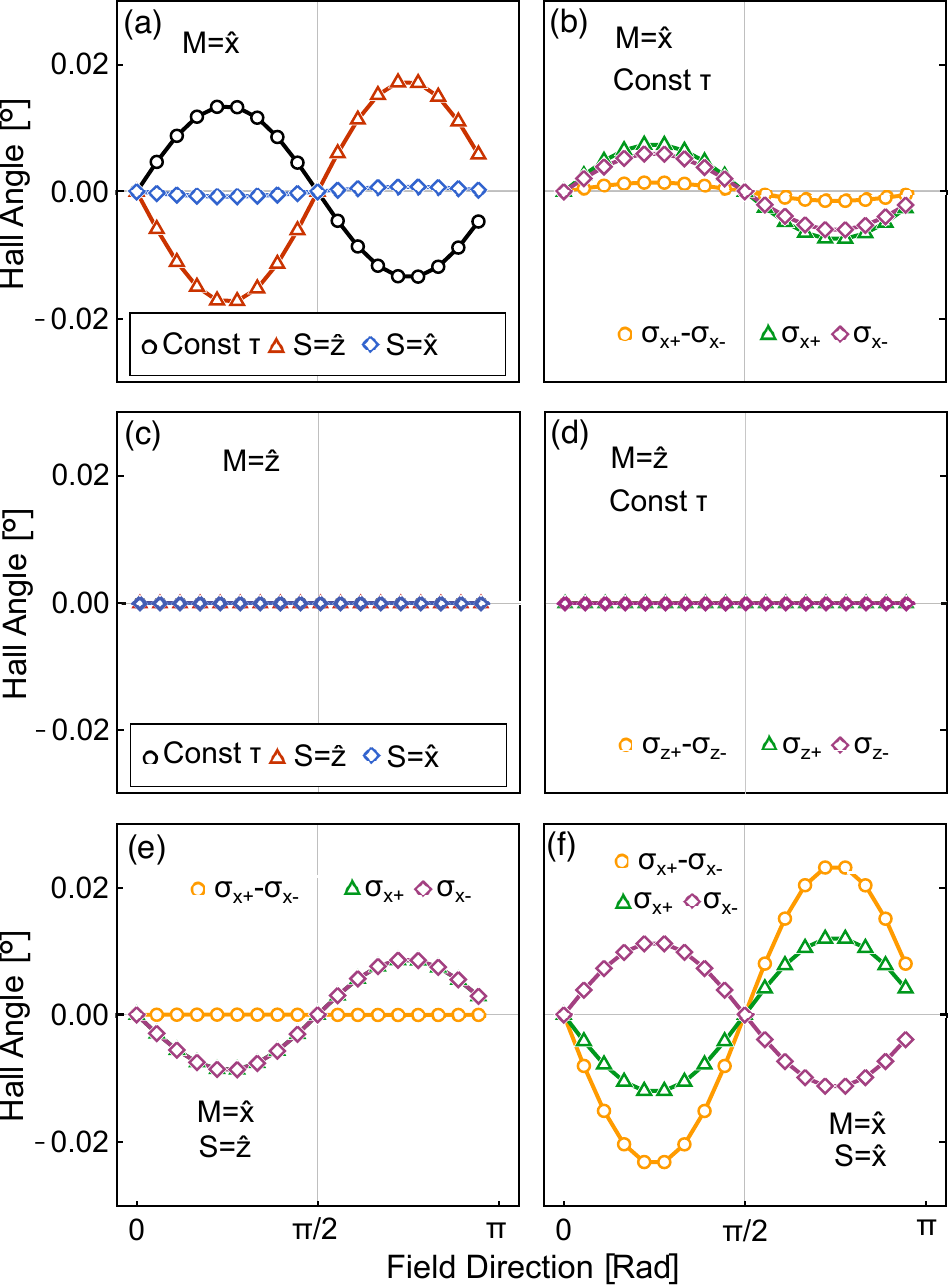} \caption{(a) The Hall angle as a function of the direction of the electric
field. The magnetization of the monolayer FGT is brought to $\mathbf{M}=\hat{x}$.
The dark curve is the result of constant relaxation time approximation
(RTA), whereas the dark red and light blue curves represent the cases
beyond RTA, where the spin-polarized disorders along different directions
are taken into account for the full-band scattering. (b) The spin
components of the case of RTA. (c) The Hall angles forbidden by the
symmetry captured numerically when $\mathbf{M}=\hat{z}$. (d) The
spin components of the symmetry prohibited Hall angles for the case
of RTA shown in (c). (e) The spin components of the finite Hall angles
after the ${\cal C}_{3}$ symmetry is broken by $\mathbf{M}=\hat{x}$.
Here the impurity polarization is along $\mathbf{S}=\hat{z}$. (f)
The pure spin Hall effect when the impurity spins are rotated to $\mathbf{S}=\hat{x}$.
\label{fig:gHall}}
\end{figure}
\begin{figure}
\begin{centering}
\includegraphics[width=0.95\columnwidth]{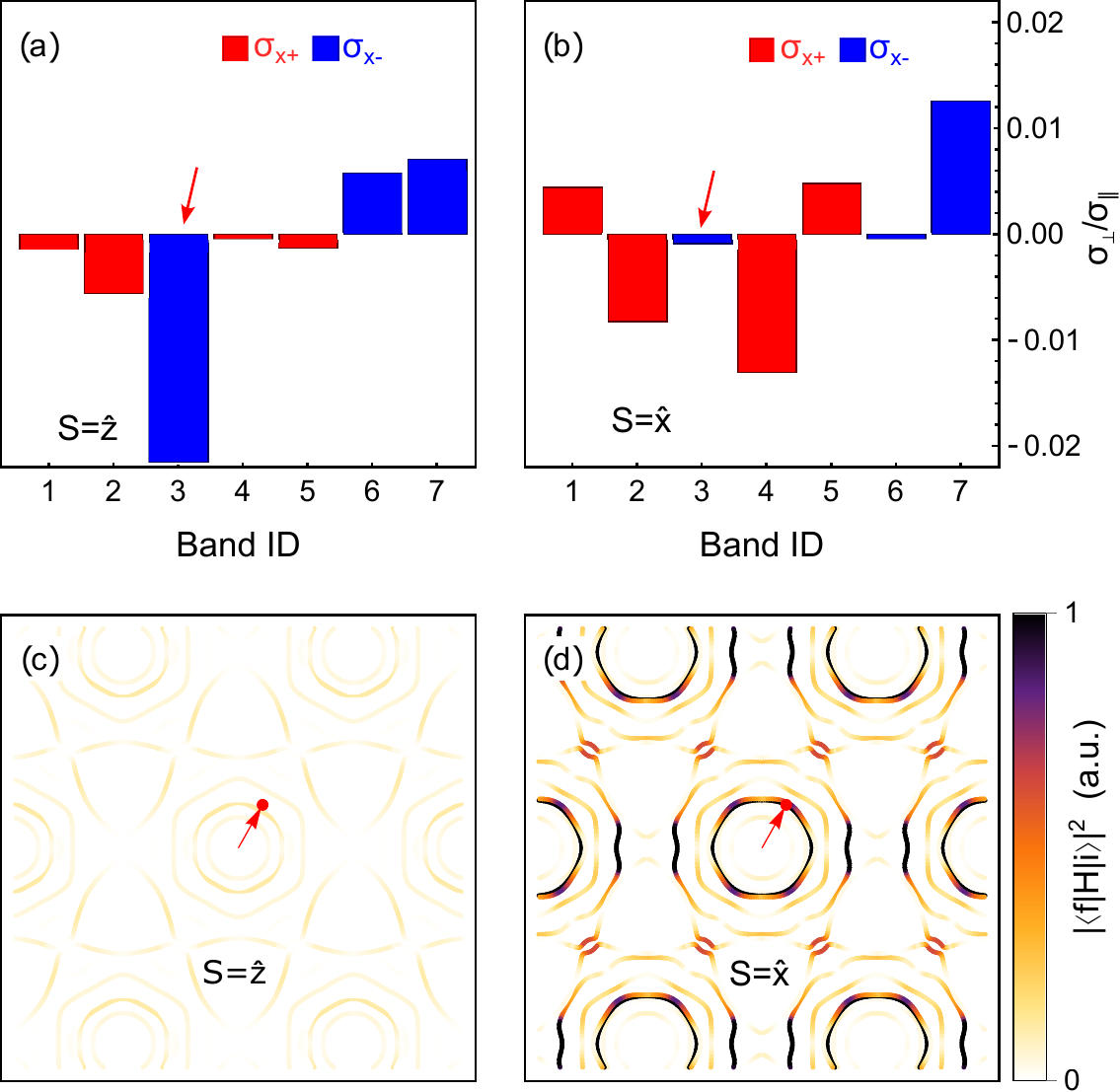}
\par\end{centering}
\centering{}\includegraphics[width=1\columnwidth]{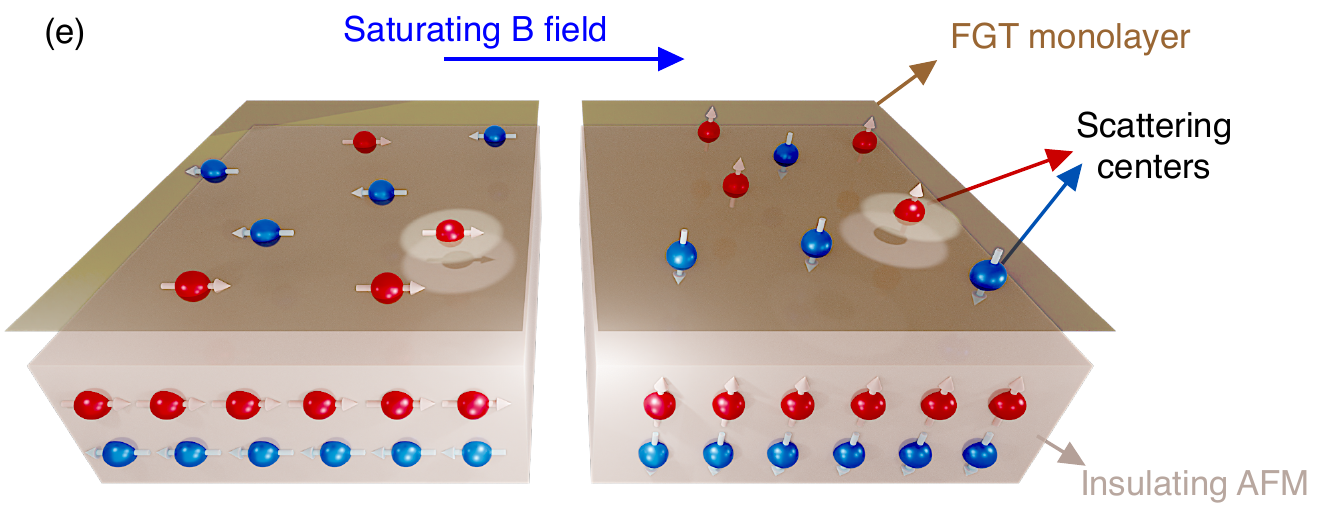}\caption{Band resolved (spin) Hall conductivity for (a) $\mathbf{S}=\hat{z}$
and (b) $\mathbf{S}=\hat{x}$. The band index counts from $\Gamma$
within the 1st Brillouin zone. (c-d) The magnitude of the scattering
matrix element from an initial state (dark red arrow) in the 3rd band
to all other states for $\mathbf{S}=\hat{z}$ and $\mathbf{S}=\hat{x}$,
respectively. (e) The scheme of a heterostructure detecting antiferromagnetic
switching. \label{fig:Mechanism}}
\end{figure}
%

Once $\{\lambda_{\mathbf{k}}^{\parallel}\}$ are obtained using Eq. \ref{eq:boltzmann:e4},
we calculate the Hall angle using $\tan\theta_{H}=\sigma_{\perp}/\sigma_{\parallel}$,
which depends only on the properties of scattering mechanisms and
the geometry of the Fermi surface\citep{ong_geometric_1991}. 
Here we illustrate the Hall angles given by the full-band Boltzmann equation
at different directions of the $\boldsymbol{{\cal E}}$ field {[}\fref{fig:gHall}(a){]}
and compare them to the RTA result (dark curve) where the details
of the scattering mechanisms are neglected. 
Clearly, the sign of the
Hall angle changes for different scattering mechanisms. 
To understand
this trend, we compare the contributions given by different spins.
For RTA, the bands with positive and negative spins almost have the
same Hall angle as shown in Fig. \ref{fig:gHall}(b). 
These two angles
sum to the dark curve in Fig. \ref{fig:gHall}(a), suggesting a net
finite extrinsic anomalous Hall effect of the electric current. 
The spin Hall angle can
then be obtained by subtracting the two spin contributions, resulting
a near-zero spin Hall angle as shown by the orange curve in Fig. \ref{fig:gHall}(b).
The finite Hall effects captured here are a consequence of the
FGT monolayer magnetization along $\hat{x}$, breaking the ${\cal C}_{3}$
rotation symmetry. 
Consistent with the discussion shown in Eq. \ref{eq:IntegralWithC3},
when the magnetization is brought to $\mathbf{M}=\hat{z}$, the ${\cal C}_{3}$
symmetry is restored and the Hall effects are strictly forbidden,
as numerically captured in Figs. \ref{fig:gHall}(c-d). 
%

Beyond RTA, the scenarios of Hall effects dramatically change once
the scattering mechanisms are turned on. When the impurity spins are
polarized along $\mathbf{S}=\hat{z}$ whereas keeping the magnetization
along $\mathbf{M}=\hat{x}$, the Hall angles for both positive and
negative spins change sign, resulting in a net negative Hall angle
when $\boldsymbol{{\cal E}}$ sweeps within $[0,\frac{\pi}{2}]$ as
shown in Fig. \ref{fig:gHall}(e). This suggests the important role
of the scattering details. More interestingly, when the scattering
centers are rotated to $\mathbf{S}=\hat{x}$, only the Hall angle
for negative spins changes sign, resulting in a finite net spin Hall
effect as shown by the orange curve in Fig. \ref{fig:gHall}(f). In
this case, the net Hall effect of the electric current vanishes, which
is consistent with the light-blue line in Fig. \ref{fig:gHall}(a). 

The transition between electric and spin Hall currents shown in Figs.
\ref{fig:gHall}(e-f) is not a consequence of symmetry. Instead, it
is determined by the details of the Fermi surface and the scattering
mechanism. Particularly, the transition is induced by the sign change
of Hall angles for the negative-spin bands denoted by the purple curves
in Figs. \ref{fig:gHall}(e-f). To understand this, contributions
to $\nicefrac{\sigma_{\perp}}{\sigma_{\parallel}}$ are resolved by
the band indices and the spins, as illustrated in Figs. \ref{fig:Mechanism}(a-b).
Here the angle of the electric field is kept at $50^{\circ}$ from
the magnetization. When $\mathbf{S}=\hat{z}$, the 3rd band (dark
red arrow, counting from $\Gamma$) on the Fermi surface contributes
a large, negative Hall angle. Such contribution is given by a band
of $|x-\rangle$, resulting in a net negative Hall angle for negative
spins. On the other hand, when $\mathbf{S}=\hat{x}$, the contribution
from the 3rd band almost vanishes {[}Fig. \ref{fig:Mechanism}(b){]}.
This can also be seen from the scattering matrix elements with a chosen
initial state in the 3rd band, as shown in Figs. \ref{fig:Mechanism}(c-d).
When $\mathbf{S}=\hat{z}$, the 3rd band is forbidden to scatter into
the van Hove singularities {[}Fig. \ref{fig:Mechanism}(c){]}. On
the other hand, such scattering is allowed when $\mathbf{S}=\hat{x}$,
resulting in a small contribution to the transport and the Hall angle.
The overall outcome is therefore a net positive Hall angle for negative
spins, as shown by the purple curve in Fig. \ref{fig:gHall}(f) at
$50^{\circ}$. Such positive spin Hall angle almost exactly cancels
the negative Hall angle given by positive spins {[}green in Fig. \ref{fig:gHall}(f){]},
such that the net Hall angle for the electric charge becomes vanishingly
small. However, the magnitude of the spin Hall angle is maximized
{[}orange in Fig. \ref{fig:gHall}(f){]}, representing a pure spin
current along the transverse direction. 

The transition between the electric and spin Hall effects in a monolayer
FGT is a plausible way to detect the switching of an adjacent insulating
magnetic material using carrier transport. This is particularly useful
to detect the switching of an antiferromagnet (AFM) since the scattering
selection rule is only sensitive to the orientation of the spins of
the disorder, instead of the sign. Without electrons near the Fermi
surface, insulating AFMs are expected to have small damping for the
N\'eel-vector dynamics, allowing for switchings even faster than
metallic ones. However, such switching is difficult to detect, which
often involves optical imaging, spin-wave detection, or other non-trivial
instrumentation. In a heterostructure illustrated in Fig. \ref{fig:Mechanism}(e),
the FGT monolayer is weakly coupled to an insulating AFM, such that
the transport in the FGT layer experiences the sparse impurities with
opposite spins provided by the AFM layer. The scattering selection
rule survives the sign change of the disorder spins and is therefore
robust against the surface roughness. Although the heterostructure
illustrated in Fig. \ref{fig:Mechanism}(e) suggests a layered AFM
structure, collinear AFMs with other spin structures can also be detected
via this mechanism. This may provide a convenient playground to investigate
the spintronics of insulating antiferromagnets in general. 
\begin{acknowledgments}
\emph{Acknowledgments:} This paper is based upon work supported by
the National Science Foundation (US) under Grant No. ECCS-2151809.
This work used Bridges-2 at Pittsburgh Supercomputing Center through
allocation PHY230018 from the Advanced Cyberinfrastructure Coordination
Ecosystem: Services \& Support (ACCESS) program, which is supported
by National Science Foundation (US) grants \#2138259, \#2138286, \#2138307,
\#2137603, and \#2138296.
\end{acknowledgments}

\providecommand{\noopsort}[1]{}\providecommand{\singleletter}[1]{#1}%

\end{document}